\documentclass[conference]{IEEEtran-lex}

\RequirePackage{snapshot}

\usepackage{graphicx}
\usepackage{color}
\usepackage{underscore}
\usepackage{balance}

\usepackage{subcaption}
\usepackage{authblk}
\usepackage{amsmath}
\usepackage{float}
\usepackage{cuted}

\usepackage{hyperref}
\hypersetup{
    colorlinks=true,
    linkcolor=blue,
    filecolor=magenta,      
    urlcolor=cyan,
    citecolor=magenta,
}
\newcommand{\secref}[1]{\S\ref{sec:#1}}
\newcommand{\figref}[1]{Fig.~\ref{fig:#1}}

\newcommand{\miles}{511,638}
\newcommand{\frames}{7.1 billion}
\newcommand{\faceframes}{2.2 billion} % ratio should be 1.1/3.5
\newcommand{\subjects}{122}
\newcommand{\days}{15,610}
\newcommand{\months}{37}
\newcommand{\vehicles}{29}
\usepackage[printwatermark]{xwatermark}
\usepackage{xcolor}
\usepackage{fancyhdr}

\fancypagestyle{firststyle}
{
  \fancyhf{}
  
   \fancyfoot[L]{
     \footnotesize
     * Corresponding author: Lex Fridman (\texttt{fridman@mit.edu}). Linda Angell and Sean Seaman are affiliated with Touchstone Evaluations, Inc. All other authors are affiliated with Massachusetts
     Institute of Technology (MIT).
   }
}

\begin{document}

\bstctlcite{IEEEexample:BSTcontrol}

% \title{MIT Autonomous Driving Study}
% \title{MIT Autonomous Driving Study:\\{\huge Large-Scale Dataset for Deep Learning Based Analysis of Driver Interaction with
%   Vehicle Automation}}
% \title{MIT Autonomous Driving Study:\\
%   \textcolor[gray]{0.5}{Large-Scale Deep Learning Based Analysis of Driver Interaction with Vehicle Automation}
% }

% \title{MIT Autonomous Vehicle Technology Study:\\
%   \textcolor[gray]{0.5}{Large-Scale Deep Learning Based Analysis of Driver Behavior and Interaction with Automation}}

\title{MIT Advanced Vehicle Technology Study:\\
  \textcolor[gray]{0.5}{Large-Scale Naturalistic Driving Study of\\Driver Behavior and Interaction with Automation}}

\author{
  Lex Fridman$^*$,
  Daniel E. Brown,
  Michael Glazer,
  William Angell,
  Spencer Dodd,  
  Benedikt Jenik,\\
  Jack Terwilliger,
  Aleksandr Patsekin,
  Julia Kindelsberger,
  Li Ding,
  Sean Seaman,
  Alea Mehler,\\
  Andrew Sipperley,
  Anthony Pettinato,
  Bobbie Seppelt,
  Linda Angell,
  Bruce Mehler,
  Bryan Reimer
}

%\affil{Massachusetts Institute of Technology (MIT)}

% \newcommand{\authspace}{\hspace{1in}}
% \author{
% Lex Fridman \authspace\and
% Daniel E. Brown \authspace\and
% Michael Glazer \authspace\and
% William Angell \authspace\and
% Spencer Dodd \authspace\and
% Hillary Abraham \authspace\and
% %Benedikt Jenik \authspace\and
% Bobbie Seppelt \authspace\and
% Bruce Mehler \authspace\and
% Bryan Reimer
% }

% \author{Lex Fridman}
% \author{Daniel E. Brown}
% \author{Michael Glazer}
% \author{William Angell}
% \author{Spencer Dodd}
% \author{Hillary Abraham}
% \author{Benedikt Jenik}
% \author{Bobbie Seppelt}
% \author{Bruce Mehler}
% \author{Bryan Reimer}
% \affil{Massachusetts Institute of Technology (MIT)}

% \author{\IEEEauthorblockN{Michael Shell}
% \and
% \IEEEauthorblockN{Homer Simpson}
% \and
% \IEEEauthorblockN{James Kirk\\ and Montgomery Scott}
% \IEEEauthorblockA{Starfleet Academy\\
% San Francisco, California 96678--2391\\
% Telephone: (800) 555--1212\\
% Fax: (888) 555--1212}}

\maketitle
%\setlength{\footskip}{100pt}
%\setlength{\headsep}{.36in}

% \begin{figure}[H]
%   \centering
%   \includegraphics[width=\textwidth]{images/cars/cars3.jpg}
%   %\caption{Illustration of the two ``arguing machines'' under investigation in this work.}
%   \label{fig:cars}
% \end{figure}

\begin{abstract}
  Today, and possibly for a long time to come, the full driving task is too complex an activity to be fully formalized as a sensing-acting robotics system that can be explicitly solved through model-based and learning-based approaches in order to achieve full unconstrained vehicle autonomy. Localization, mapping, scene perception, vehicle control, trajectory optimization, and higher-level planning decisions associated with autonomous vehicle development remain full of open challenges. This is especially true for unconstrained, real-world operation where the margin of allowable error is extremely small and the number of edge-cases is extremely large. Until these problems are solved, human beings will remain an integral part of the driving task, monitoring the AI system as it performs anywhere from just over 0\% to just under 100\% of the driving. The governing objectives of the MIT Advanced Vehicle Technology (MIT-AVT) study are to (1) undertake large-scale real-world driving data collection that includes high-definition video to fuel the development of deep learning based internal and external perception systems, (2) gain a holistic understanding of how human beings interact with vehicle automation technology by integrating video data with vehicle state data, driver characteristics, mental models, and self-reported experiences with technology, and (3) identify how technology and other factors related to automation adoption and use can be improved in ways that save lives. In pursuing these objectives, we have instrumented 23 Tesla Model S and Model X vehicles, 2 Volvo S90 vehicles, 2 Range Rover Evoque, and 2 Cadillac CT6 vehicles for both long-term (over a year per driver) and medium term (one month per driver) naturalistic driving data collection. Furthermore, we are continually developing new methods for analysis of the massive-scale dataset collected from the instrumented vehicle fleet. The recorded data streams include IMU, GPS, CAN messages, and high-definition video streams of the driver face, the driver cabin, the forward roadway, and the instrument cluster (on select vehicles). The study is on-going and growing. To date, we have \subjects{} participants, \days{} days of participation, \miles{} miles, and \frames{} video frames. This paper presents the design of the study, the data collection hardware, the processing of the data, and the computer vision algorithms currently being used to extract actionable knowledge from the data.
\end{abstract}

\thispagestyle{firststyle}
\setlength{\footskip}{10pt}

\begin{strip}
  \centering  \includegraphics[width=0.97\textwidth]{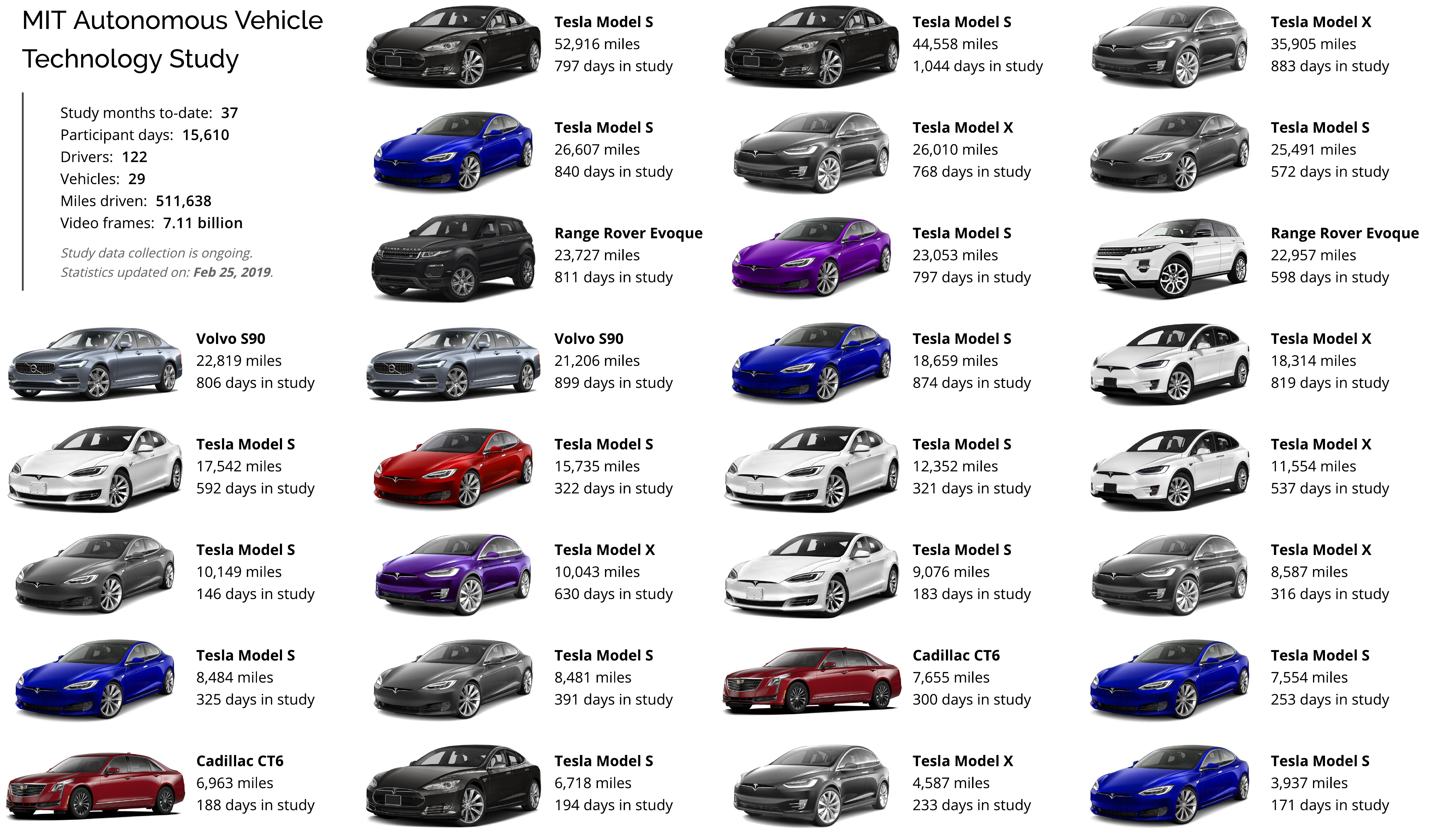}
  \captionof{figure}{Dataset statistics for the MIT-AVT study as a whole and for the individual vehicles in the study.}
  \label{fig:cars}
\end{strip}

\newcommand{\cvfig}[2]{%
  \begin{subfigure}[t]{\columnwidth}
    \includegraphics[width=\textwidth]{images/computer-vision/final-#1.jpg}%
    \caption{#2}
    \label{fig:cv-#1}
  \end{subfigure}\hspace{0.06in}
}
\newcommand{\cvspace}{\vspace{0.1in}}

\begin{figure}[H]
  \centering
  \cvfig{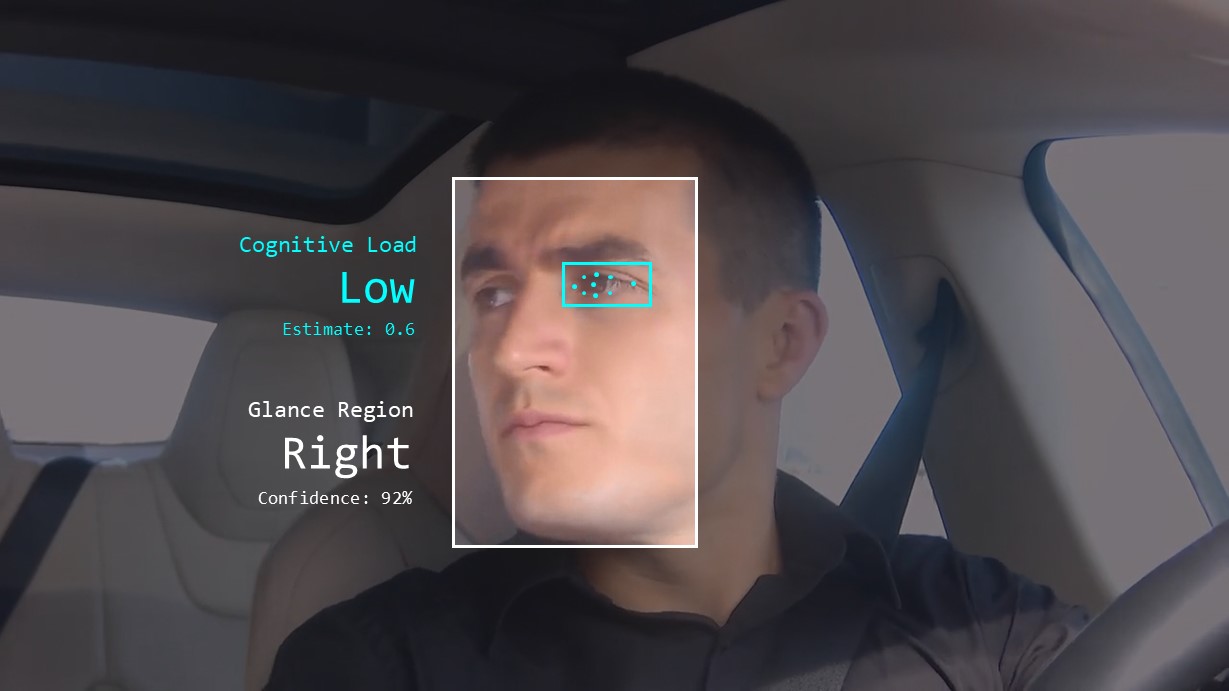}{Face Camera for Driver State.}
  \\\cvspace
  \cvfig{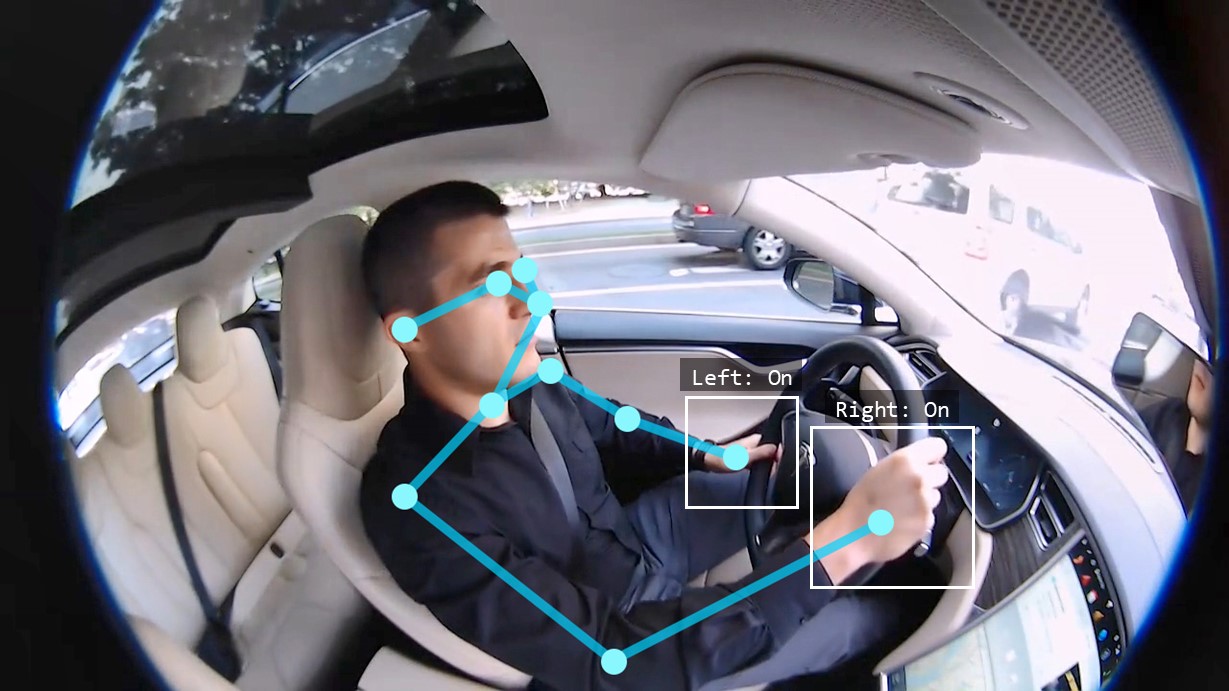}{Driver Cabin Camera for Driver Body Position.}
  \\\cvspace
  \cvfig{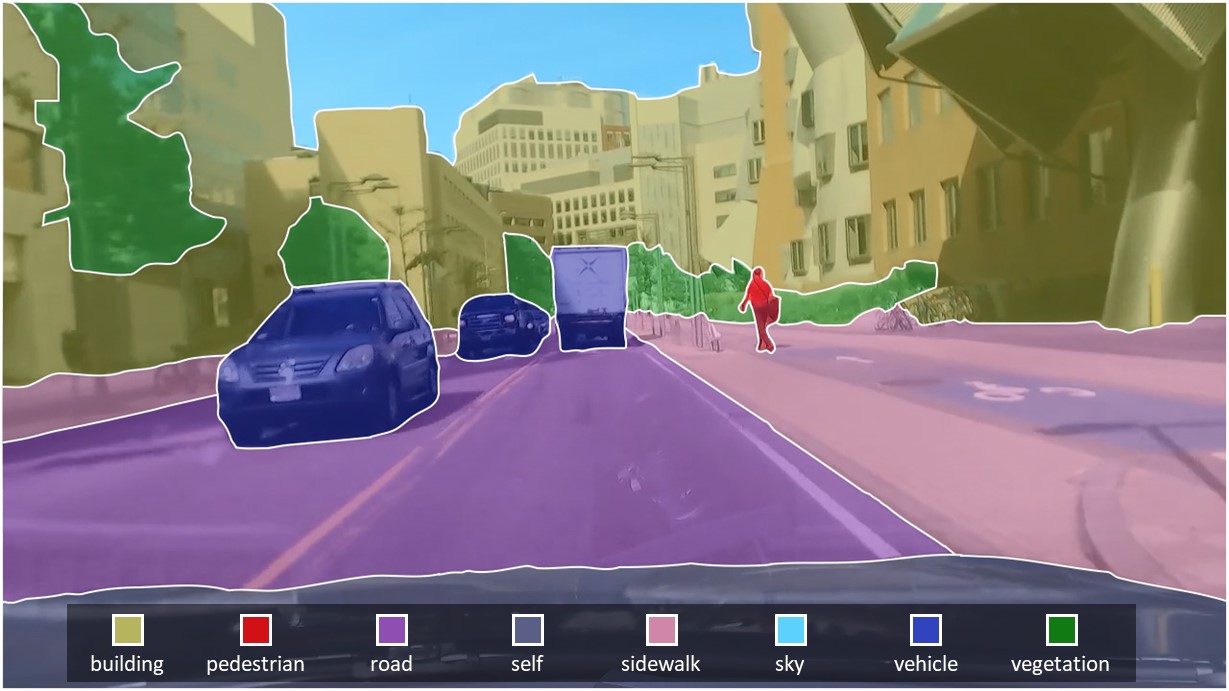}{Forward-Facing Camera for Driving Scene Perception.}
  \\\cvspace
  \cvfig{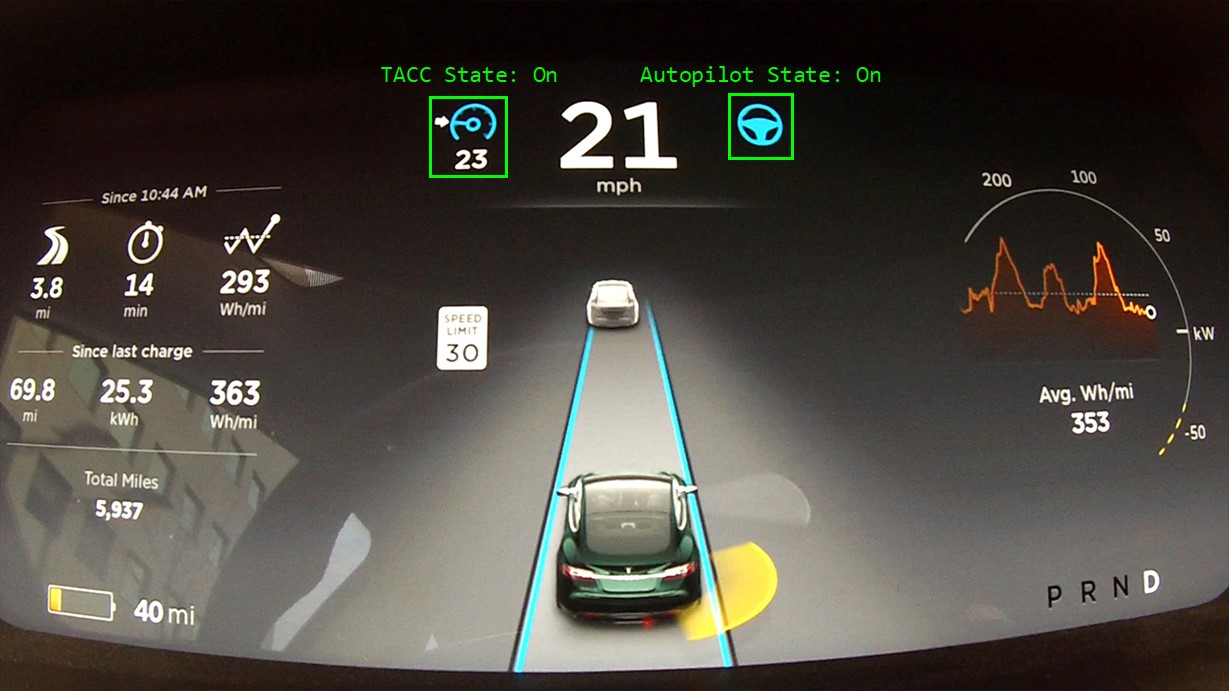}{Instrument Cluster Camera for Vehicle State.}
  \caption{Video frames from MIT-AVT cameras and visualization of computer vision tasks performed for each.}
  \label{fig:cv}
\end{figure}

\begin{figure*}[ht!]
  \centering
  {%
    \setlength{\fboxsep}{0pt}%
    \setlength{\fboxrule}{0.3pt}%
    \fbox{\includegraphics[width=\textwidth]{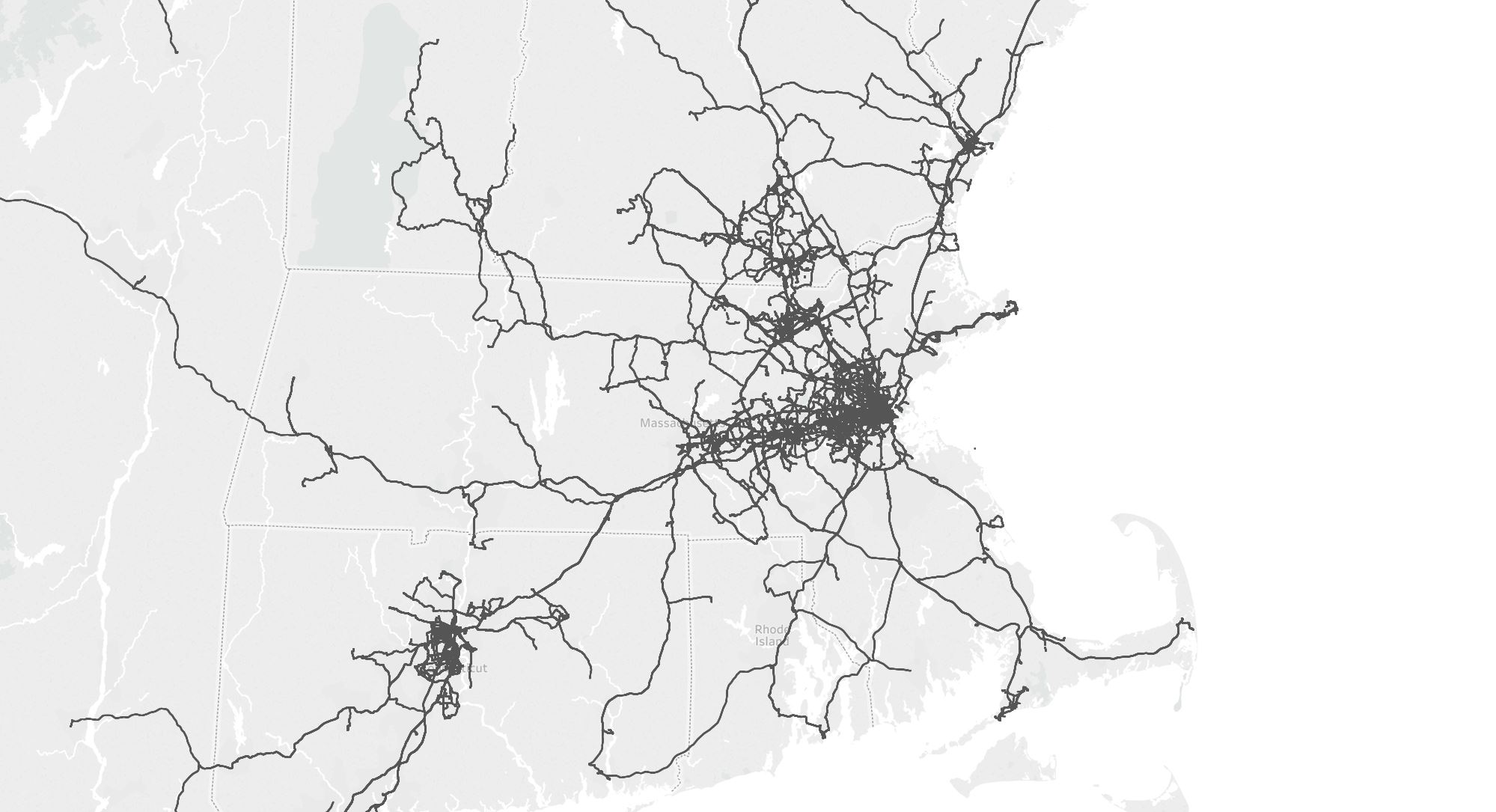}}
    \caption{Visualization of GPS points for trips in the MIT-AVT dataset local to the New England area. The full
      dataset contains trips that span over the entire continental United States.}
    \label{fig:gps-all}
  }%
\end{figure*}

\thispagestyle{plain}
\pagestyle{plain}
\setlength{\footskip}{40pt}

\section{Introduction}\label{sec:introduction}

The idea that human beings are poor drivers is well-documented in popular culture
\cite{davies2015look,vanderbilt2009traffic}. While this idea is often over-dramatized, there is some truth to it in that
we're at times distracted, drowsy, drunk, drugged, and irrational decision makers \cite{world2015global}. However, this
does not mean it is easy to design and build a perception-control system that drives better than the average human
driver. The 2007 DARPA Urban Challenge \cite{buehler2009darpa} was a landmark achievement in robotics, when 6 of the 11
autonomous vehicles in the finals successfully navigated an urban environment to reach the finish line, with the first
place finisher traveling at an average speed of 15 mph. The success of this competition led many to declare the fully
autonomous driving task a ``solved problem'', one with only a few remaining messy details to be resolved by automakers
as part of delivering a commercial product. Today, over ten years later, the problems of localization, mapping, scene
perception, vehicle control, trajectory optimization, and higher-level planning decisions associated with autonomous
vehicle development remain full of open challenges that have yet to be fully solved by systems incorporated into a
production platforms (e.g. offered for sale) for even a restricted operational space. The testing of prototype vehicles
with a human supervisor responsible for taking control during periods where the AI system is ``unsure'' or unable to
safely proceed remains the norm \cite{dixit2016autonomous,favaro2017examining}.

The belief underlying the MIT Advanced Vehicle Technology (MIT-AVT) study is that the DARPA Urban Challenge was only a
first step down a long road toward developing autonomous vehicle systems. The Urban Challenge had no people
participating in the scenario except the professional drivers controlling the other 30 cars on the road that day. The
authors believe that the current real-world challenge is one that has the human being as an integral part of every
aspect of the system. This challenge is made especially difficult due to the immense variability inherent to the driving
task due to the following factors:

\begin{itemize}
\item The underlying uncertainty of human behavior as represented by every type of social interaction and conflict
  resolution between vehicles, pedestrians, and cyclists.
\item The variability between driver styles, experience, and other characteristics that contribute to their
  understanding, trust, and use of automation.
\item The complexities and edge-cases of the scene perception and understanding problem.
\item The underactuated nature of the control problem \cite{tedrake2014underactuated} for every human-in-the-loop
  mechanical system in the car: from the driver interaction with the steering wheel to the tires contacting the road
  surface.
\item The expected and unexpected limitation of and imperfections in the sensors.
\item The reliance on software with all the challenges inherent to software-based systems: bugs, vulnerabilities, and the
  constantly changing feature set from minor and major version updates.
\item The need for a human driver to recognize, acknowledge, and be prepared to take control and adapt when system
  failure necessitates human control of the vehicle in order to resolve a potentially dangerous situation.
\item The environmental conditions (i.e., weather, light conditions) that have a major impact on both the low-level
  perception and control tasks, as well as the high-level interaction dynamics among the people that take part in the
  interaction.
\item Societal and individual tolerances to human and machine error.
\end{itemize}

As human beings, we naturally take for granted how much intelligence, in the robotics sense of the word, is required to
successfully attain enough situation awareness and understanding \cite{endsley1995out} to navigate through a world full
of predictably irrational human beings moving about in cars, on bikes, and on foot. It may be decades before the majority of
cars on the road are fully autonomous. During this time, the human is likely to remain the critical decision maker either
as the driver or as the supervisor of the AI system doing the driving.

In this context, Human-Centered Artificial Intelligence (HCAI) is an area of computer science, robotics, and experience
design that aims to achieve a deeper integration between human and artificial intelligence. It is likely that HCAI will
play a critical role in the formation of technologies (algorithms, sensors, interfaces, and interaction paradigms) that
support the driver's role in monitoring the AI system as it performs anywhere from just over 0\% to just under 100\% of
the basic driving and higher order object and event detection tasks. 

The MIT Advanced Vehicle Technology (MIT-AVT) study seeks to collect and analyze large-scale naturalistic data of
semi-autonomous driving in order to better characterize the state of current technology use, to extract insight on how
automation-enabled technologies impact human-machine interaction across a range of environments, and to understand how
we design shared autonomy systems that save lives as we transition from manual control to full autonomy in the coming
decades. The effort is motivated by the need to better characterize and understand how drivers are engaging with
advanced vehicle technology \cite{reimer2014driver}. The goal is to propose, design, and build systems grounded in this
understanding, so that shared autonomy between human and vehicle AI does not lead to a series of unintended consequences
\cite{barry2011toomuch}.

``Naturalistic driving'' refers to driving that is not constrained by strict experimental design and a ``naturalistic
driving study'' (NDS) is generally a type of study that systematically collects video, audio, vehicle telemetry, and
other sensor data that captures various aspects of driving for long periods of time, ranging from multiple days
to multiple months and even years. The term NDS is applied to studies in which data are acquired under conditions that
closely align with the natural conditions under which drivers typically drive ``in the wild.''  Often, a driver's own
vehicle is instrumented (as unobtrusively as possible) and each driver is asked to continue using their vehicle as
they ordinarily would. Data is collected throughout periods of use. Further, use is unconstrained by any
structured experimental design.  The purpose is to provide a record of natural behavior that is as unaffected by the
measurement process as possible. This contrasts with on-road experiments that are conducted in similarly instrumented
vehicles, but in which experimenters are present in the vehicle, and ask drivers to carry out specific tasks at specific
times on specific roads using specific technology systems in the vehicle.

The MIT-AVT study is a new generation of NDS that aims to discover insights and understanding of real-world interaction
between human drivers and autonomous driving technology. Our goal is to derive insight from large-scale naturalistic
data being collected through the project to aid in the design, development and delivery of new vehicle systems, inform
insurance providers of the changing market for safety, and educate governments and other non-governmental stakeholders
on how automation is being used in the wild.

This paper outlines the methodology and underlying principles governing the design and operation of the MIT-AVT study
vehicle instrumentation, data collection, and the use of deep learning methods for automated analysis of human
behavior. These guiding principles can be summarized as follows:

\begin{itemize}
\item \textbf{Autonomy at All Levels:} We seek to study and understand human behavior and interaction with every form of
  advanced vehicle technology that assists the driver through first sensing the external environment and the driver
  cabin, and then either controlling the vehicle or communicating with the driver based on the perceived state of the
  world. These technologies include everything from automated emergency braking systems that can take control in rare
  moments of imminent danger to semi-autonomous driving technology (e.g., Autopilot) that can help control the lateral
  and longitudinal movements of the vehicle continuously for long periods of driving on well-marked roadways (e.g., highways).
\item \textbf{Beyond Epochs and Manual Annotation:} Successful large-scale naturalistic driving studies of the past in
  the United States \cite{neale2005overview, dingus2006100, klauer2006impact, campbell2012shrp,victor2015analysis} and
  in Europe \cite{benmimoun2011incident} focused analysis on crash and near-crash epochs. Epochs were detected using
  traditional signal processing of vehicle kinematics. The extraction of driver state from video was done primarily with
  manual annotation. These approaches, by their nature, left the vast remainder of driving data unprocessed and
  un-analyzed. In contrast to this, the MIT-AVT study seeks to analyze the ``long-tail'' of shared-autonomy from both
  the human and machine perspectives. The ``long-tail'' is the part of data that is outside of short, easily-detectable
  epochs. It is, for example, the data capturing moment-to-moment allocation of glance over long stretches of driving
  (hundreds of hours in MIT-AVT) when the vehicle is driving itself. Analyzing the long-tail data requires processing
  billions of high-definition video frames with state-of-the-art computer vision algorithms multiple times as we learn
  both what to look for and how to interpret what we find. At the same time, despite the focus on deep learning based
  analysis of large-scale data, the more traditional NDS analytic approaches remain valuable, including manual
  annotation, expert review of data, insight integration from technology suppliers, and contextualizing observed
  naturalistic behavior with driver characteristics, understanding, and perceptions of vehicle technology.
\item \textbf{Multiple Study Duration:} We seek understanding human behavior in semi-autonomous systems both from
  the long-term perspective of over 1 year in subject-owned vehicles and from a medium-term perspective of 1 month in
  MIT-owned vehicles. The former provides insights into use of vehicle technology over time and the latter
  provides insights about initial interactions that involve learning the limitations and capabilities of each
  subsystem in a fashion more closely aligned with a driver's experience after purchasing a new vehicle equipped with a
  suite of technology that the driver may or may not be familiar with.
\item \textbf{Multiple Analysis Modalities:} We use computer vision to extract knowledge from cameras that look at the
  driver face, driver body, and the external driving scene, but we also use GPS, IMU, and CAN bus data to add rich
  details about the context and frequency of technology use. This data is further complemented by detailed
  questionnaire and interview data that comprise driver history, exposure to various automated and non-automated
  technologies, mental model evaluation, perceptions of safety, trust, self-reported use, and enjoyment. With this
  interdisciplinary approach, the dataset allows for a holistic view of real-world advanced technology use, and
  identifies potential areas for design, policy, and educational improvements.
\end{itemize}

The key statistics about the MIT-AVT study as a whole and about the individual vehicles in the study are shown in
\figref{cars}. The key measures of the data with explanations of the measures are as follows:

\newcommand{\whisper}[1]{\textcolor[gray]{0.5}{\small(#1)}\vspace{0.02in}}
\begin{itemize}
\item \textbf{Study months to-date:} \months{}\\
  \whisper{Number of months the study has been actively running with vehicles on the road.}
\item \textbf{Participant days:} \days{}\\
  \whisper{Number of days of active data logger recording across all vehicles in the study.}
\item \textbf{Drivers:} \subjects{}\\
  \whisper{Number of consented drivers across all vehicles in the study.}
\item \textbf{Vehicles:} \vehicles{}\\
  \whisper{Number of vehicles in the study.}
\item \textbf{Miles driven:} \miles{}\\
  \whisper{Number of miles driven.}
\item \textbf{Video frames:} \frames{}\\
  \whisper{Number of video frames recorded and processed across all cameras and vehicles in the study.}
\end{itemize}

Latest dataset statistics can be obtained at \url{http://hcai.mit.edu/avt} (see \secref{conclusion}).
%The growth of the dataset over time is shown in \figref{growth-and-calendar}.
Data collection is actively on-going. \figref{gps-all}
shows GPS traces for trips in the dataset local to the New England Area.

%%%%%%%%%%%%%%%%%%%%%%%%%%%%%%%%%%%%%%%%%%%%

\subsection{Naturalistic Driving Studies}

The focus of the MIT-AVT study is to gather naturalistic driving data and to build on the work and lessons-learned of
the earlier generation of NDS studies carried out over the first decade of the 21st century
\cite{neale2005overview,dingus2006100,klauer2006impact,campbell2012shrp,victor2015analysis}. These previous studies
aimed to understand human behavior right before and right after moments of crashes and near-crashes as marked by periods
of sudden deceleration. The second Strategic Highway Research Program (SHRP2) is the best known and largest scale of
these studies \cite{campbell2012shrp}.

In contrast to SHRP-2 and other first-generation NDS efforts, the MIT-AVT study aims to be the standard for the next
generation of NDS programs where the focus is on large-scale computer vision based analysis of human behavior. Manually
annotating specific epochs of driving, as the prior studies have done, is no longer sufficient for understanding the
complexities of human behavior in the context of autonomous vehicle technology (i.e., driver glance or body position
over thousands of miles of Autopilot use). For example, one of many metrics that are important to understanding driver
behavior is moment-by-moment detection of glance region \cite{fridman2016driver,fridman2016owl} (see
\secref{deep-learning-driving}). In order to accurately extract this metric from the \faceframes{} frames of face video
without the use of computer vision would require an immense investment in manual annotation, assuming the availability
of an efficient annotation tool that is specifically designed for the manual glance region annotation task and can
leverage distributed, online, crowdsourcing of the annotation task. The development of such a tool is a technical
challenge that may take several years of continuous research and development \cite{russell2008labelme}, which may
eclipse the cost human annotation hours. If this was the only metric of interest, perhaps such a significant investment
would be justifiable and feasible. However, glance region is only one of many metrics of interest, and in terms of
manual annotation cost, is one of the least expensive. Another example is driving scene segmentation, which for
\faceframes{} frames would require an incredible investment \cite{cordts2016cityscapes}. For this reason, automatic or
semi-automatic extraction of information from raw video is of paramount importance and is at the core of the motivation,
design, research, and operation of MIT-AVT.

The fundamental belief underlying our approach to NDS is that only by looking at the entirety of the data (with
algorithms that reveal human behavior and situation characteristics) can we begin to learn which parts to ``zoom in''
on: which triggers and markers will lead to analysis that is representative of system performance and human behavior in
the data \cite{knipling2015naturalistic, knipling2017crash, fridman2017arguing, shankar2008analysis,
  kalra2016driving}. Furthermore, each new insight extracted from the data may completely change our understanding of
where in the data we should look. For this reason, we believe understanding how humans and autonomous vehicles interact
requires a much larger temporal window than an epoch of a few seconds or even minutes around a particular event. It
requires looking at the long-tail of naturalistic driving that has up until now been largely ignored. It requires
looking at entire trips and the strategies through which humans engage the automation: when, where, and for how long it
is turned on, when and where it is turned off, when control is exchanged, and many other questions. Processing this huge
volume of data necessitates an entirely different approach to data analysis. We perform the automated aspect of the
knowledge extraction process by using deep learning based computer vision approaches for driver state detection, driver
body pose estimation, driving scene segmentation, and vehicle state detection from the instrument cluster video as shown
in \figref{cv} and discussed in \secref{software}. This work describes the methodology of data collection that enabled
the deep learning analysis. Individual analysis effort are part of future follow-on work. The result of using deep
learning based automated annotation is that MIT-AVT can analyze the long-tail of driving in the context of shared
autonomy, which in turn, permits the integration of complex observed interactions with the human's perception of their
experience. This innovative interdisciplinary approach to analysis of NDS datasets in their entirety offers a unique
opportunity to evaluate situation understanding of human-computer interaction in the context of automated driving.

\subsection{Datasets for Application of Deep Learning}

Deep learning \cite{goodfellow2016deep} can be defined in two ways: (1) a branch of machine learning that uses neural
networks that have many layers or (2) a branch of machine learning that seeks to form hierarchies of data representation
with minimum input from a human being on the actual composition of the hierarchy. The latter definition is one that
reveals the key characteristic of deep learning that is important for our work, which is the ability of automated
representation learning to use large-scale data to generalize robustly over real-world edge cases that arise in any
in-the-wild application of machine learning: occlusion, lighting, perspective, scale, inter-class variation, intra-class
variation, etc. \cite{hartley2003multiple}.

In order to leverage the power of deep learning for extracting human behavior from raw video, large-scale annotated
datasets are required. Deep neural networks trained on these datasets can then be used for their learned representation
and then fine-tuned for the particular application in the driving context. ImageNet \cite{deng2009imagenet} is an image
dataset based on WordNet \cite{miller1990introduction} where 100,000 synonym sets (or ``synsets'') each define a unique
meaningful concept. The goal for ImageNet is to have 1000 annotated images for each of the 100,000 synsets. Currently it
has 21,841 synsets with images and a total of 14,197,122 images. This dataset is commonly used to train neural network
for image classification and object detection tasks \cite{he2016deep}. The best performing networks are highlighted as
part of the annual ImageNet Large Scale Visual Recognition Competition (ILSVRC) \cite{russakovsky2015imagenet}. In this
work, the terms ``machine learning,'' ``deep learning,'' ``neural networks,'' and ``computer vision'' are often used
interchangeably. This is due to the fact that the current state-of-the-art for most automated knowledge extraction tasks
are dominated by learning-based approaches that rely on one of many variants of deep neural network
architectures. Examples of other popular datasets leveraged in the development of algorithms for large-scale analysis of
driver behavior in our dataset include:

\begin{itemize}
\item \textbf{COCO} \cite{lin2014microsoft}: Microsoft Common Objects in Context (COCO) dataset is a large-scale dataset
  that addresses the object detection task in scene understanding under two perspectives: detecting non-iconic views of
  objects, and the precise 2D localization of objects. The first task usually refers to object localization, which uses
  bounding boxes to denote the presence of objects. The second task refers to instance segmentation, for which the
  precise masks of objects are also needed. The whole dataset features over 200,000 images labeled within 80 object
  categories. Successful methods~\cite{he2016deep,he2017mask,dai2017deformable} jointly model the two tasks together and
  simultaneously output bounding boxes and masks of objects.

\item \textbf{KITTI} \cite{Geiger2013IJRR, Geiger2012CVPR}: KITTI driving dataset develops challenging benchmarks for
  stereo vision, optical flow, visual odometry / SLAM and 3D object detection, captured by driving around in both rural
  areas and highways of Karlsruhe (a mid-size city in Germany). In total, there are 6 hours of traffic scenarios
  recorded at 10-100 Hz using a variety of sensor modalities such as high-resolution color and grayscale stereo cameras,
  a Velodyne 3D laser scanner and a high-precision GPS/IMU inertial navigation system. In addition, \cite{Menze2015CVPR}
  also propose ground truth for 3D scene flow estimation by collecting 400 highly dynamic scenes from the raw dataset
  and augmenting them with semi-dense scene flow ground truth.

\item \textbf{Cityscapes} \cite{Cordts2015TheCD}: The Cityscapes dataset focuses on semantic understanding of urban
  street scenes. It offers a large, diverse set of stereo video sequences recorded in streets from 50 different cities
  with pixel-level and instance-level semantic labeling. There are 5,000 fully segmented images with pixel-level
  annotations and an additional 20,000 partially segmented images with coarse annotations. Its two benchmark challenges
  have led to the development of many successful approaches for semantic segmentation
  \cite{zhao2017pyramid,wang2017understanding} and instance segmentation \cite{he2017mask,liu2017iccv}.

\item \textbf{CamVid} \cite{brostow2009semantic}: Cambridge-driving Labeled Video Database (CamVid) is the first dataset
  with frame-wise semantic labels in videos captured from the perspective of a driving automobile. The dataset provides
  ground truth labels that associate each pixel with one of 32 semantic classes. Manually specified per-pixel semantic
  segmentation of over 700 images total enables research on topics such as pedestrian
  detection~\cite{tian2015pedestrian}, and label propagation~\cite{badrinarayanan2010label}.
\end{itemize}

\subsection{Automotive Applications of  Deep Learning}\label{sec:deep-learning-driving}

Design of perception and control systems in the driving domain have benefited significantly from learning-based
approaches that leverage large-scale data collection and annotation in order to construct models that generalize over
the edge cases of real-world operation. Leveraging the release large-scale annotated driving datasets
\cite{Geiger2013IJRR,Cordts2015TheCD}, automotive deep learning research aims to address detection, estimation,
prediction, labeling, generation, control, and planning tasks. As shown in \figref{cv}, specific tasks have been defined
such as fine-grained face recognition, body pose estimation, semantic scene perception, and driving state
prediction. Current efforts are briefly summarized as follows:

\begin{itemize}
\item \textbf{Fine-grained Face Recognition}: Beyond classic face recognition studies, fine-grained face recognition
  focuses on understanding human behavior toward face perception, such as facial expression
  recognition~\cite{liu2014facial,yu2015image}, eye gaze detection~\cite{hoffman2000distinct,wisniewska2014robust}. In
  the driving context, \cite{fridman2017can,vicente2015driver} explore the predictive power of driver
  glances. \cite{gao2014detecting,abdic2016driver} use facial expression to detect emotional stress for driving safety
  and the driving experience.
	
\item \textbf{Body Pose Estimation}: Work on human body pose expands the performance, capabilities, and experience of
  many real-world applications in robotics and action recognition. Successful approaches vary from using depth
  images~\cite{shotton2013real}, via deep neural networks~\cite{toshev2014deeppose}, or with both convolutional networks
  and graphical models~\cite{tompson2014joint}. Specifically for driving, \cite{sadigh2014data} use driver pose, which
  is represented by skeleton data including positions of wrist, elbow, and shoulder joints, to model human driving
  behavior. \cite{mbouna2013visual} cast visual analysis of eye state and head pose for driver alertness monitoring.
	
\item \textbf{Semantic Scene Perception}: Understanding the scene from 2D images has long been a challenging task in
  computer vision, which often refers to semantic image segmentation. By taking advantage of large scale datasets like
  Places~\cite{zhou2014learning}, Cityscapes~\cite{Cordts2015TheCD}, many
  approaches~\cite{zhao2017pyramid,wang2017understanding} manage to get state-of-the-art results with powerful deep learning
  techniques. As a result, precise driving scene perception~\cite{xu2017end,bojarski2016end} for self-driving cars is
  now actively studied in both academia and industry.
	
\item \textbf{Driving State Prediction}: Vehicle state is usually considered as a direct illustration of human decision
  in driving, which is also the goal for autonomous driving. In terms of machine learning, it serves as the ground truth
  for various tasks from different perspectives such as driving behavior~\cite{sadigh2014data} and steering
  commands~\cite{xu2017end, bojarski2016end}.	
\end{itemize}

Many aspects of driver assistance, driver experience, and vehicle performance are increasingly being automated with
learning-based approaches as representative datasets for these tasks are released to the broad research community. The
MIT-AVT study aims to be the source of many such datasets that help train neural network architectures that provide
current and future robust solutions for many modular and integrated subtasks of semi-autonomous and fully-autonomous driving.

\section{MIT-AVT Study Structure and Goals}

%\lexdo{Include an overview of the study similar to the introduction, and tying together the following sections.}

The governing principle underlying the design of all hardware, low-level software, and higher-level data processing
performed in the MIT-AVT study is: continual, relentless innovation, while maintaining backward compatibility. From the
beginning, we chose to operate at the cutting-edge of data collection, processing, and analysis approaches. This meant
trying a lot of different approaches and developing completely new ones: from sensor selection and hardware design
described in \secref{hardware} to the robust time-critical recording system and the highly sophisticated data pipeline
described in \secref{software}. It's a philosophy that allowed us to scale quickly and find new solutions at every level
of the system stack.

% \subsection{Study Time Scales: One Month and Multiple Years}\label{sec:time-scales}

% \lexdo{Describe the design of month-long and 1+ year long naturalistic driving study motivation.}

% \subsection{Large-Scale Computer Vision: From Pixels to Understanding}

% \lexdo{Describe what machines are very good and somewhat good at detecting automatically.}

% \subsection{Manual, Semi-Automated, and Fully-Automated Annotation}\label{sec:annotation}

% \lexdo{Describe what machines are not good at detecting and the role of human annotation for secondary tasks and for
%   assisting the annotation for the previous task.}
  
% \lexdo{Motivate annotation for behavioral understanding and annotation for computer vision}

\subsection{Participation Considerations and Recruitment}\label{sec:recruitment}

As previously noted, the medium duration (one month long) NDS is conducted using MIT-owned vehicles, while the long
duration (over 1 year) NDS is conducted in subject-owned vehicles. Participants are divided into primary and secondary
drivers, all of whom, in order to take part in the study, must formally agree to the terms detailed in an informed
consent form approved by an institutional review board (IRB). Primary drivers in the long NDS (usually the most frequent
driver of the vehicle or the car owner) must be willing to provide permission to install the data acquisition equipment
in the vehicle, warning labels on windows to advise non-consented passengers and drivers of the ongoing data collection,
and coordinate with project staff for system maintenance and data retrieval. Recruitment is conducted through flyers,
social networks, forums, online referrals, and word of mouth. Primary drivers are compensated for their time involvement
in vehicle instrumentation, system maintenance appointments, data retrieval, and completing questionnaires.

To be accepted as a primary driver in an MIT-owned vehicle fleet requires that potential subjects' daily commutes
include time on specific highways, a willingness to use a study vehicle for a period of approximately four weeks as the
subject's primary commuting vehicle, signing an IRB approved informed consent form, passing a Criminal Offender Record
Information (CORI) check and driving record review by MIT's Security and Emergency Management Office, participating in a
training protocol that covers both basic and advanced vehicle features, and completing a series of questionnaires and
interviews prior to and after their naturalistic driving experience. High-level overviews of the training protocol,
questionnaire, and interview strategies can be found in \secref{training-one-month} and
\secref{qual-approaches-one-month}, respectively.

\subsection{Training Conditions for One Month NDS}\label{sec:training-one-month}

Participants in the medium duration (one month long) NDS are provided with introductions to the fleet vehicles in the
form of an approximately 1.5 hour long training session. This session is intended to introduce drivers to the physical
characteristics of the vehicle, and provide a sufficient understanding of vehicle features in order to support safe use
of advanced technologies. Participants are provided with a study overview by a researcher and presented with
manufacturer produced videos or information packets on one or more of the basic and advanced features available in the
vehicle. After this initial introduction to systems outside of the vehicle, participants are seated in the vehicle and
given a guided overview of the vehicle layout and settings (e.g. seat / mirror adjustments, touchscreen menu
layout). Participant's phones are paired with the vehicle, and they are given the opportunity to practice several voice
commands (e.g. placing a phone call, entering a destination). Next, more detailed overviews are provided on the
function, activation, and use of the following features:

\begin{itemize}
\item Adaptive Cruise Control (ACC)
\item Pilot Assist (in the Volvo)
\item Super Cruice (in the Cadillac)
\item Forward Alert Warning / City Safety (in the Volvo)
\item Automatic Emergency Braking
\item Lane Departure Warning (LDW)
\item Lane Keep Assist (LKA)
\item Blind Spot Monitor
\end{itemize}

Following this stationary in-vehicle training, participants are provided with an on-road training drive on a multi-lane
highway. This highway driving session lasts a minimum of 30 minutes to allow for practical exposure to the systems in
real world setting. During the training drive participants are encouraged to utilize the researcher and ask questions
when testing out the systems. Participants are encouraged to customize vehicle settings to their preferences and to
develop sufficient familiarity to support the ability to choose to use or not use certain systems for the duration of
their one month period of vehicle use.

\subsection{Qualitative Approaches for One Month NDS}\label{sec:qual-approaches-one-month}

Self-report data collection methods are kept as unobtrusive to participation in the study as possible, while still
capturing the richness of driver's experience with the vehicle and various systems, their thoughts on the technology
after participating, and barriers toward their intentions to adopt or discard automation moving forward. Self-report
data in the medium duration (one month long) NDS is captured using three questionnaire batteries and one semi-structured
interview. Self-report data is collected prior to and after the naturalistic portion of the experiment; at no point are
participants asked to complete questionnaires or interviews while they are in possession of the vehicle.

The questionnaire batteries are deployed in three stages. The first occurs when a subject signs the consent form and
completes the background check paperwork. The first questionnaire collects basic demographics and information on driving
history, driving style, exposure to various advanced and established in-vehicle technologies, and general trust in
technology. A second questionnaire is completed immediately following the training protocol outlined in
\secref{training-one-month}, and captures participants' high level mental models, initial impressions, and reported
trust in select vehicle technologies. The third and final questionnaire is completed at the end of the driver's
one-month naturalistic driving period. This questionnaire assesses reported trust in select technologies, perceptions of
safety, high- and detailed-level understanding of systems, and desire for having in their own future vehicles such
systems as experienced during the NDS period and with hypothetical improvements. Many questions in the second and third
questionnaires are identical, allowing analysis to explore how exposure to systems and experiential learning impact
concepts such as trust and understanding of technologies.

A semi-structured interview is conducted in person between a research associate and the study participant at the end of
the one-month naturalistic driving period, and lasts approximately 30-60 minutes. It consists of predefined questions
focusing on initial reactions to the vehicle, experience during the training drive, how training affected their
understanding of the technologies, and driver perceptions of the technologies.
 
\subsection{Competitors Collaborate: Consortium Model}

Naturalistic driving data and automated deep learning based interpretation of that data gives insights, suggestions, and
well-grounded scenarios as to the path forward for safe and effective integration of artificial intelligence into modern
and future vehicle systems. The raw data and the high-level understanding of human behavior and system performance in
such autonomous vehicle technology is of interest to:

\begin{itemize}
\item Car companies (both established and newly formed)
\item Automotive parts suppliers 
\item Insurance companies
\item Technology companies
\item Government agencies
\item Academic and research organization
\end{itemize}

When the path forward is full of uncertainty, risks, potentially costly misaligned investments, and paradigm shifts, open
innovation provides more value than closed competition. At this moment in time, autonomous vehicle technology is a space
where competitors win by collaborating, sharing high-level insights and large-scale, real-world data. 

High-level measures such as system use and system performance can be used to inform the design, development and
validation of future vehicle systems. Basic driver behavior with and without technology use can fuel basic research on
driver understanding, use characteristics, and decision models while aiding in the actuation of risk in the insurance
market. Video recording inside and out of the vehicle can be used to develop perception, control, planning, driver
sensing, and driver assistance systems. As such, the data collected in the MIT-AVT study can be leveraged for a range of
quantitative and qualitative efforts. Members of the Advanced Vehicle Technology consortium \cite{misc2016avt} are
collaborating to support the acquisition of data through the MIT-AVT study, development of new data processing
approaches, and selected analysis. Full members of the consortia have rights to data access for proprietary or other
internal use purposes. Several members of the effort are actively involved in independent research (with and without MIT
involvement) using MIT-AVT study data.

% \section{Vehicles and Autonomous Vehicle Technology in the Study}

% \lexdo{Add a beautiful image of all the systems in our study.}

% \lexdo{Write all the following sections.}

% \subsection{Sensors}

% \subsection{General Systems}

% \subsubsection{Automated Emergency Braking}

% \subsubsection{Warnings}

% \paragraph{Forward Collision Warning (FCW)}

% \paragraph{Lane Departure Warning (LDW)}

% \paragraph{Blind Spot Detection}

% \subsection{Adaptive Cruise Control}

% \subsection{Intelligent Systems that Need Computer Vision}

% \subsection{Tesla Autopilot}

% \subsection{Volvo Pilot Assist II}

% \subsection{Range Rover Evoque Lane Keep Assist}

% \subsection{Cadillac Super Cruise}

\section{Hardware: Data Logging and Real-Time Processing}\label{sec:hardware}

\begin{figure}
  \centering
  \includegraphics[width=\columnwidth]{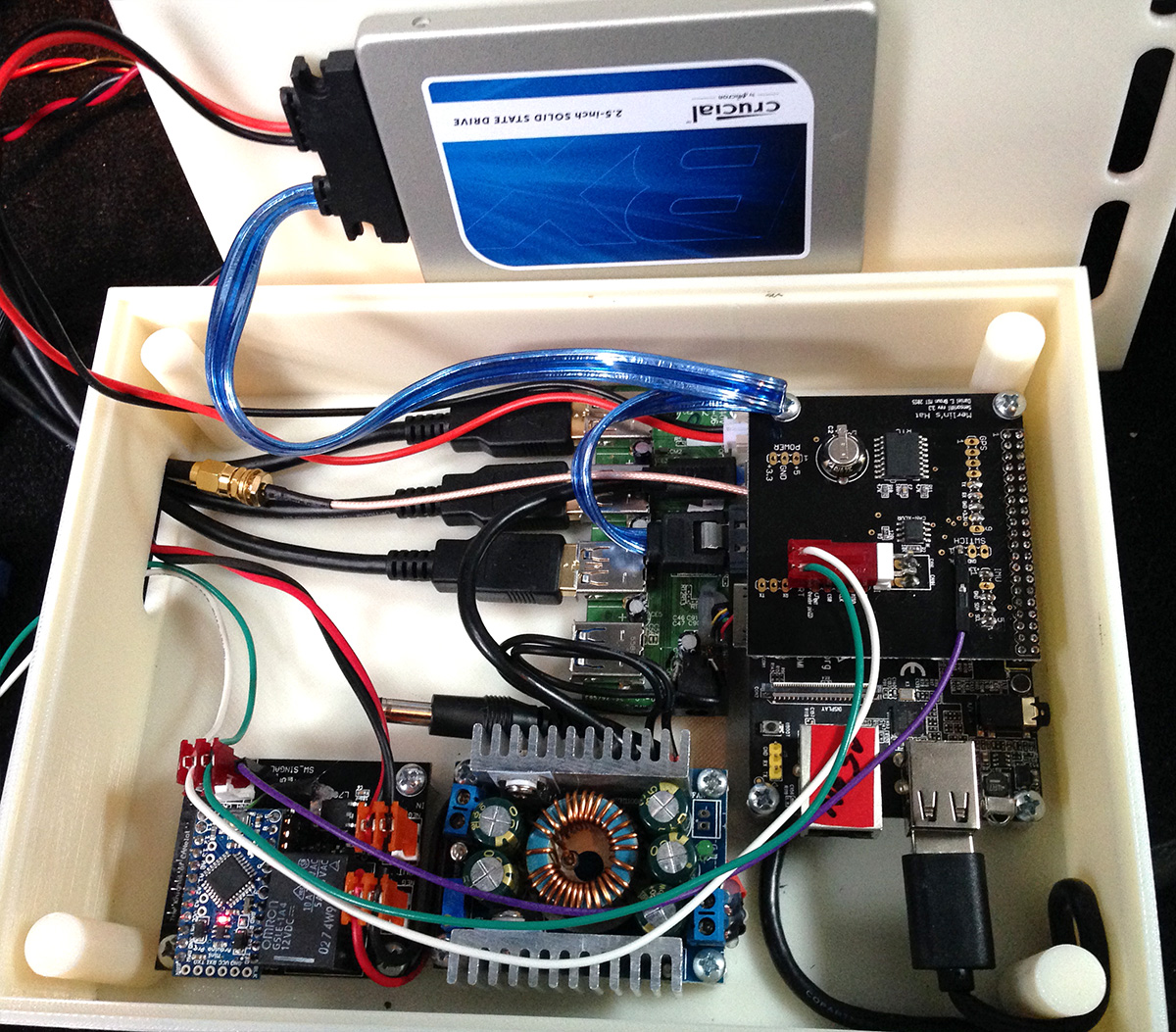}
  \caption{Final prototype version of RIDER enclosed by 3D printed case. From top to bottom, clockwise, attached to the
    top of the case is external storage in the form of a 1 terabyte solid state hard drive. The USB cameras connect via
    a USB hub shown in the center. To the right of the USB hub, Banana Pi covered by the black SensorHAT with CAN
    transceiver, GPS, IMU, and real time clock. Bottom center, buck converter for stepping down vehicle battery voltage
    from 12-13.8 volts to 5 volts for all compute systems. Lower left, Knights of CANelot CAN controlled power board.}
  \label{fig:rider}
\end{figure}

The backbone of a successful naturalistic driving study is the hardware and low-level software that performs the data
collection. In the MIT-AVT study, that role is served by a system named RIDER (Real-time Intelligent Driving Environment
Recording system) as shown in \figref{rider}. RIDER was designed and continuously developed to satisfy the following
goals and requirements:

\begin{enumerate}
\item \textbf{Timestamped Asynchronous Sensor Recording:} Record all sensors and data streams in a way that each sample
  of data (no matter its frequency or data source) is timestamped using a centralized, reliable time-keeper. In other
  words, data has to be timestamped in a way that allows perfect synchronization of multiple data streams in
  post-processing \cite{fridman2016automated}.
\item \textbf{High-Definition Video:} Capture and record 3 to 6 cameras at 720p (2.1 megapixels) resolution. The
  selection of camera positions, resolution, and compression was one of the most essential design decisions of the
  entire study. See \secref{cameras} for discussion of how this selection was made.
\item \textbf{CAN Bus:} Collect vehicle telemetry from the Controller Area Network (CAN) bus(es) of the vehicle
  \cite{li2008design}. Each vehicle has different ports and bus utilization policies, with little information made
  publicly available about the mapping of message ID's and the message content. Raw CAN messages must be recorded such
  that the essential information is contained within those messages even if at the time of collection those messages
  cannot be decoded.
\item \textbf{Remote Cellular Connectivity:} Low-bandwidth, infrequent communication of system status via a cellular
  connection in order to detect when RIDER system malfunction occurs.
\item \textbf{Discrete and Elegant Appearance:} Parts of the system that are visible from inside or outside the car
  should have a small form-factor and have visual design characteristics that do not detract from the overall appearance
  of the vehicle or have an impact on the overall driving experience.
\item \textbf{Camera Mounting is Robust but Removable:} Mounting must be consistent, reliable, and removable designed
  specifically for each vehicle's interior  physical characteristics.
\end{enumerate}

RIDER components include a real-time-clock, GPS, IMU, and the ability to record up to 6 cameras at 720p resolution,
remote cellular connectivity. The developed system employs the use of common components tailored to suit its needs
achieving a scalable ultra low cost, accurate, extendable and robust data recording platform.
 
To keep the electronics and stored data secure, RIDER is placed within in the trunk away from the elements and possible
disturbances from passengers. Power and CAN data cables are run from the OBD-II or diagnostic port to the trunk into
RIDER. USB cables for cameras are also run from each camera location into the trunk. All data and power cables are
secured and hidden beneath interior trim panels.

\subsection{Power Management System}

The power systems for RIDER has many constraints: it demanded flexibility to transfer into different vehicles and draw
minimal power when not in use as to not drain the primary vehicle battery. The power system consists of a
main smart CAN monitoring section and a buck converter. When active and logging data, RIDER draws less than 8
watts of power. When in standby, RIDER's quiescent current draw  is less than 1/10th of a watt.
 
\begin{figure}
  \centering
  \includegraphics[width=\columnwidth]{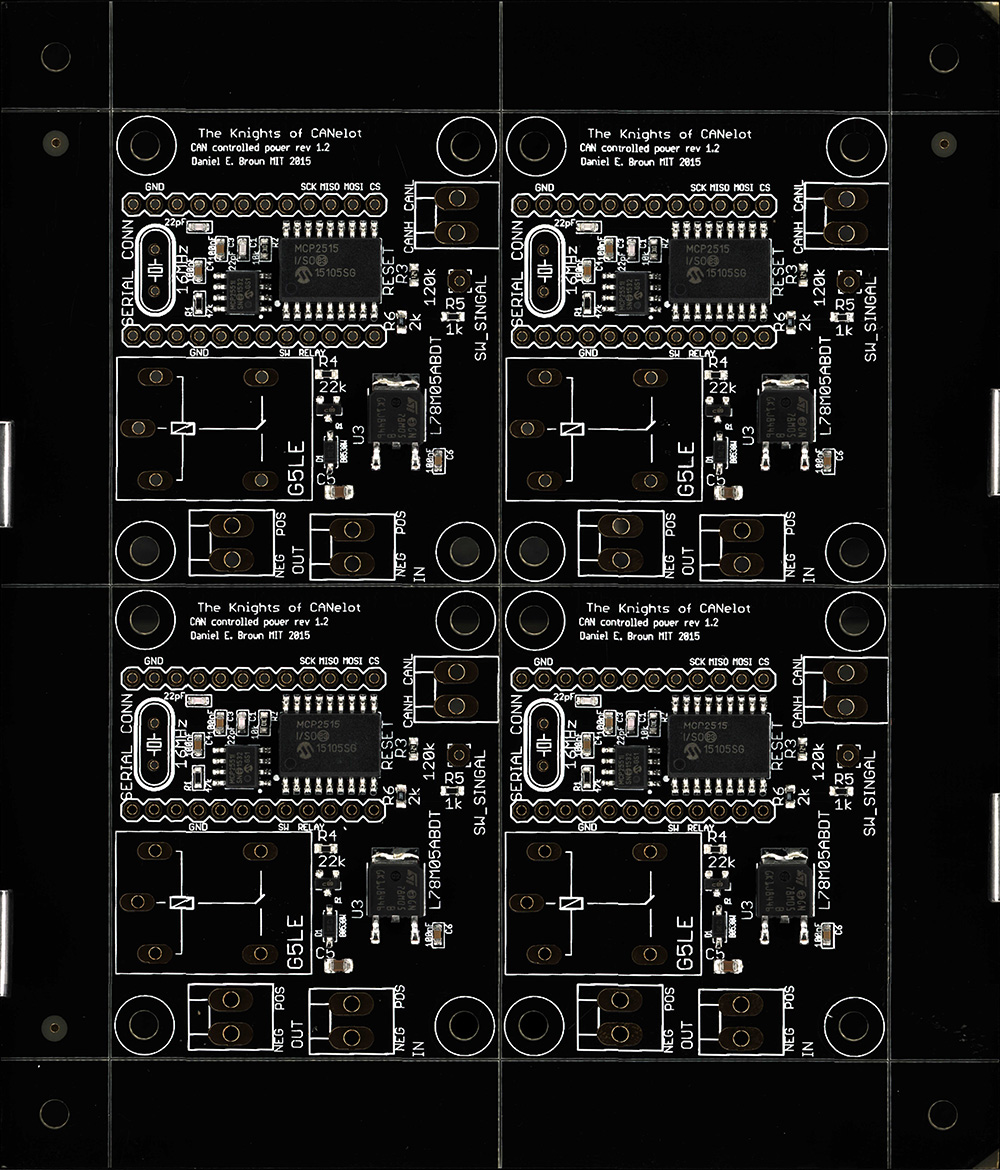}
  \caption{Knights of CANelot, CAN controlled power board. Power board mid-assembly showing populated CAN controller,
    transceiver, and power regulation. Also shown, unpopulated positions for the power relay, microcontroller,
    oscillator and connectors.}
  \label{fig:canelot-board}
\end{figure}

\begin{figure}
  \centering
  \includegraphics[width=\columnwidth]{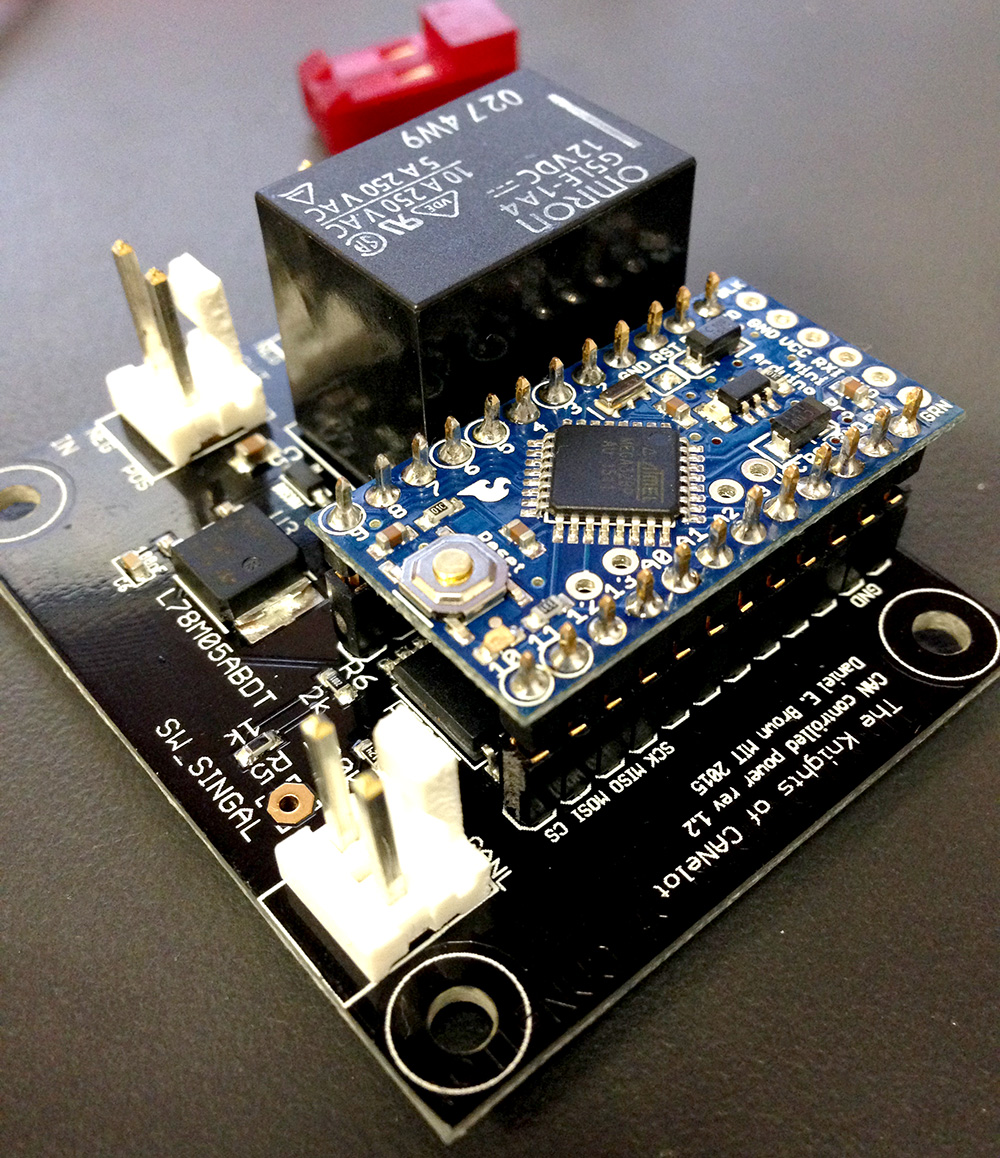}
  \caption{Fully assembled Knights of CANelot board, showing populated microcontroller, power relay, CAN and power connections.}
  \label{fig:canelot-built}
\end{figure}

The Knights of CANelot (see \figref{canelot-board} and \figref{canelot-built}) is a CAN controlled power board that
contains a microchip MCP2515 CAN controller and MCP2551 CAN transceiver, along with an Atmega328p microcontroller to
monitor CAN bus traffic. By default when powered this microcontroller places itself into sleep and does not allow power
to enter the system by way of a switching relay. When the CAN controller detects a specific predefined CAN message
indicating the vehicle CANbus is active, the microcontroller is sent an interrupt by the CAN controller waking up the
microcontroller from sleep and triggering the relay to power the primary buck converter. This begins the booting
sequence to the rest of the system. When the vehicle shuts off and the CANbus within the car enters into a sleep state,
a signal is sent via the Knights of CANelot microcontroller to gracefully stop all video and data recording, shutdown
the compute system, disconnect main power then enter sleep mode once again.

\subsection{Computing Platform and Sensors}

A single board computer was chosen for this application for its wide variety of I/O options, small form factor and ease of
development. We chose to work with the Banana Pi Pro with the follow sensors and specifications:

\begin{itemize} 
\item 1GHz ARM Cortex-A7 processor, 1GB of RAM
\item Expandable GPIO ports for IMU/GPS/CAN
\item Native onboard SATA
\item Professionally manufactured daughter board for sensor integration
\item ARM processor features onboard CAN controller for vehicle telemetry data collection
\item Maxim Integrated DS3231 real-time clock for accurate timekeeping/time-stamping +/-2 ppm accuracy
\item Texas Instruments SN65HVD230 CAN transceiver 
\item 9 degrees-of freedom inertial measurement unit (STMicro L3GD20H(gyro), LSM303D(accelerometer/compass))
\item GlobalTop MTK3339 GPS unit, 6 channel, DGPS capability accurate within 5 meters
\item Huawei E397Bu-501 4G LTE USB module
\item USB 3.0 4-port hub, powered
\item 1TB/2TB solid state hard drive
\end{itemize}
 
\subsection{Cameras}\label{sec:cameras}

Three or four Logitech C920 webcams record at a resolution of 1280x720 at 30 frames per second within the car. Two of
these cameras have been modified to accept standard CS type lens mount for adaptability within the car for either face
or body pose orientation. The third camera is the standard webcam that is mounted on the windshield for a forward road
perspective. Occasionally a fourth camera is placed within the instrument cluster to capture information unavailable on
the CANbus. These cameras also contain microphones for audio capture and recording. Custom mounts were designed for
specialty placement within the vehicle.
 
Most single board computers like our Banana Pi lack the required computational ability to encode and compress more than
one raw HD video stream. The Logitech C920 camera provides the ability to off-load compression from the compute platform
and instead takes place directly on the camera. This configuration allows for possibility of up to 6 cameras in a single
RIDER installation.
 
% [insert c920 w/case image]

% insert industrial cameras

% A customized Linux kernel was developed specifically for RIDER based on the hardware and system requirements listed
% above. The filesystem is stored on a replaceable micro SD card on-board the Banana Pi. The next section outlines and
% describes the software running within RIDER and the management of the data after collection.

\section{Software: Data Pipeline and Deep Learning Model Training}\label{sec:software}

\begin{figure*}
  \centering
  \includegraphics[width=\textwidth]{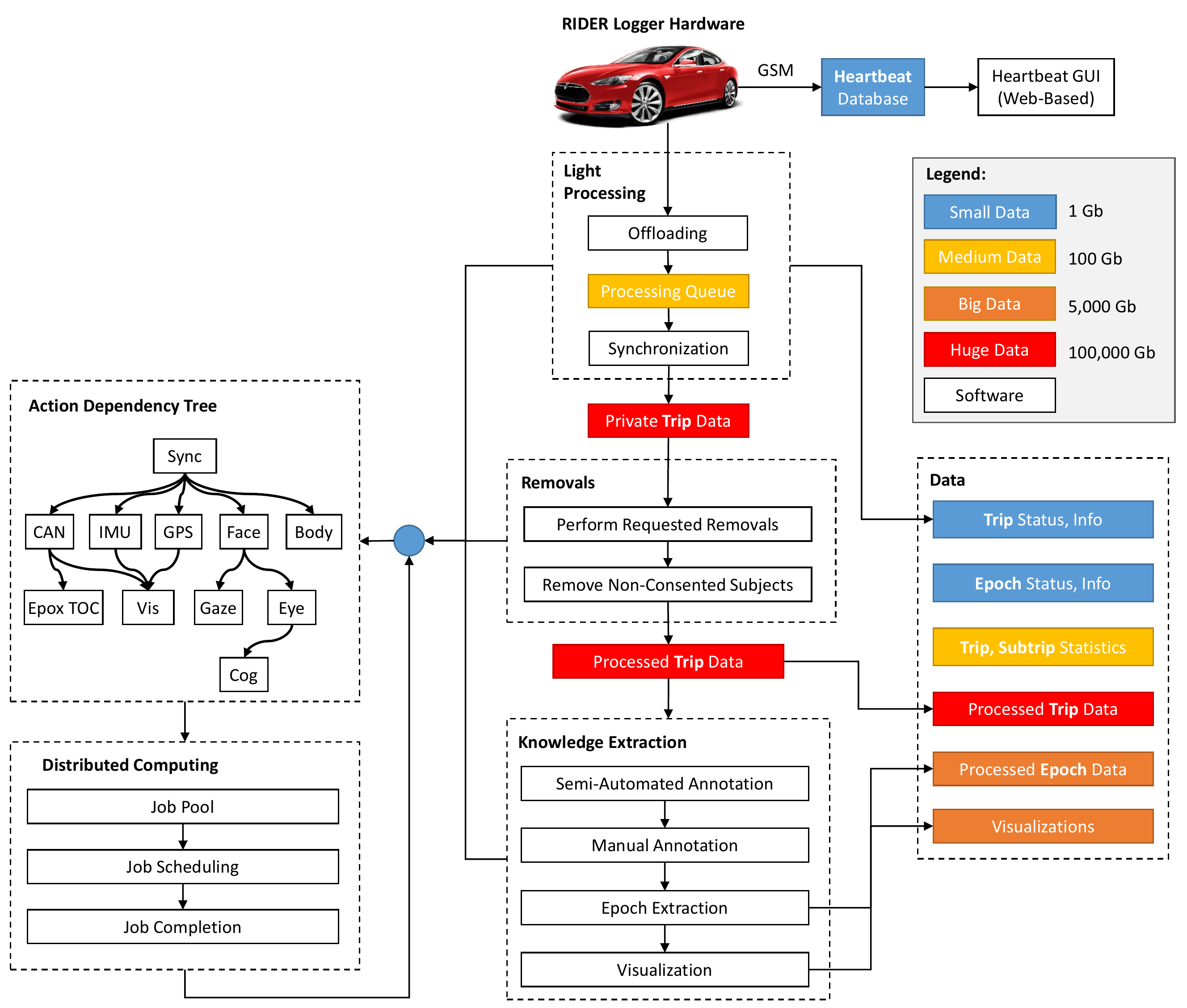}
  \caption{The MIT-AVT data pipeline, showing the process of offloading, cleaning, synchronizing, and extracting knowledge
    from data. On the left is the dependency-constrained, asynchronous, distributed computing framework. In the middle
    is the sequence of high level procedures that perform several levels of knowledge extraction. On the right are broad
    categories of data produced by the pipeline, organized by size.}
  \label{fig:pipeline}
\end{figure*}

Building on the robust, reliable, and flexible hardware architecture of RIDER is a vast software framework that handles
the recording of raw sensory data and takes that data through many steps across thousands of GPU-enabled compute cores
to the extracted knowledge and insights about human behavior in the context of autonomous vehicle
technologies. \figref{pipeline} shows the journey from raw timestamped sensor data to actionable knowledge. The
high-level steps are (1) data cleaning and synchronization, (2) automated or semi-automated data annotation,
context interpretation, and knowledge extraction, and (3) aggregate analysis and visualization.

This section will discuss the data pipeline (\figref{pipeline}), which includes software implemented on RIDER boxes that
enables data streaming and recording. In addition, the software that is used to offload and process the data on a
central server will be discussed. The operational requirement of software operating on RIDER boxes are as follows:

\begin{enumerate}
\item Power on whenever the vehicle is turned on
\item Create a trip directory on an external solid state drive
\item Redirect all data streams into timestamped trip files
\item Log and transmit metadata to the lab in real time 
\item Power down after the vehicle is turned off
\end{enumerate}

%\lexdo{goals of synchronization process if we include it?}

\subsection{Microcontroller}

The microcontroller on the Knights of CANelot power management board runs a small C program that is responsible for
powering the RIDER system in sync with the vehicle. By default, this microcontroller is in a sleep state, awaiting a
specfic CAN message. By listening to the vehicle's CANbus, this program can recognize when CAN message for a specific
signal begins, which signifies the car has turned on. If this signal is observed, the C program then connects the
vehicle's power to the rest of the system, starting the data collection. When the specified message ends, meaning the
car is off, the microcontroller sends a signal to the Banana Pi to close all files and shutdown gracefully. It then
waits 60 seconds to finally disconnect power from the rest of the system and enters its original sleep state.

\subsection{Single Board Computer}

Our single board computer, the Banana Pi, contains a 32GB SD card that stores the RIDER filesystem, software and
configuration files. The Banana Pi runs a modified Linux kernel using custom kernel modules and a tweaked Bannanian
operating system with performance and security enhancements. Performance was improved by disabling unnecessary kernel
modules and removing extraneous Linux services. Security enhancements included disabling all CAN transmission, thereby
prohibiting malicious or unintentional transmission of actuating messages to a vehicle's systems. Additional security
improvements included altering the network settings to prevent any remote connection from logging in. Specific MIT
machines were white listed to allow configuration files to be altered through a physical connection. The default system
services were also altered to run a series of locally installed programs that manage data collection whenever the system
boots.
 
\subsection{Startup Scripts}

The Banana Pi runs a series of data recording initialization bash startup scripts whenever the system boots. 
First, the on-board clock on the Pi is synchronized with a real-time clock that maintains high resolution timing
information. Modules for device communication such as UART, I2C, SPI, UVC, and CAN are then loaded to allow
interaction with incoming data streams. A monitoring script is started that shuts down the system if a specified signal is
received from the Knights of CANelot microcontroller, and an additional GSM monitoring script helps reconnect to the cellular network after losing connection. The last initialization steps are to start the python scripts Dacman and Lighthouse.

\subsection{Dacman}

Dacman represents the central data handler script that manages all data streams. It uses a configuration file called
\texttt{trip_dacman.json} that contains unique device IDs for all cameras. In addition, it contains a unique RIDER ID
associated with the RIDER box it is stored in. This configuration file also contains unique ID values for the subject,
vehicle and study this driver is associated with. Once started, Dacman creates a trip directory on the external solid
state drive named according to the date it was created using a unique naming convention:
\texttt{rider-id_date_timestamp} (e.g. \texttt{20_20160726_1469546998634990}). This trip directory contains a copy of
\texttt{trip_dacman.json}, any data related CSV (comma-separated values) files reflecting included subsystems, as well as a specifications file
called \texttt{trip_specs.json} that contains microsecond timestamps denoting the beginning and end of every subsystem
and the trip itself.

Dacman calls a manager python script for every subsystem (e.g. \texttt{audio_manager.py} or \texttt{can_manager.py}),
which makes the relevant system calls to record data. Throughout the course of the current vehicle trip, all data is
written to CSV files with timestamping information included in each row. Dacman calls two other programs written in C in
order to help generate these files: \texttt{cam2hd} for managing cameras and \texttt{dump_can} for creating CAN
files. Audio or camera data is recorded to RAW and H264 formats respectively, with an accompanying CSV denoting the
microsecond timestamp at which each frame was recorded. If any errors are encountered while Dacman is running, the
system restarts up to two times in an attempt to resolve them, and shuts down if unable to resolve them.

\subsection{Cam2HD}

\texttt{Cam2hd} is a program written in C that opens and records all camera data. It relies on V4L (Video4Linux), which
is an open source project containing a collection of camera drivers in Linux. V4L enables low level access to cameras
connected to RIDER by setting the incoming image resolution to 720p and allows the writing of raw H264 frames.

\subsection{DumpCAN}

% \lexdo{Determine if the current underscore in dumpcan is alright (the original is camelcase) and if the text emphasis
%   makes sense (note: other scripts are not currently emphasized besides cam2hd)}

\texttt{Dump_can} is another program written in C that configures and receives data from the Allwinner A20 CAN
controller. This program uses the can4linux module to produce a CSV containing all CAN data received from the connected
CANbus. In addition, it offers low level manipulation of the CAN controller. This allows \texttt{dump_can} to set listen
only mode on the can controller, which enables a heightened degree of security. By removing the need to send
acknowledgements when listening to messages on the CAN network, any possible interference with existing systems on the
CAN bus is minimized.

\subsection{Lighthouse}

Lighthouse is a python script that sends information about each trip to Homebase. Information sent includes timing
information for the trip, GPS data, power consumption, temperature and available external drive space. The interval
between communications is specified in the dacman configuration file. All communications are sent in JSON format and are
encrypted using public-key cryptography based on elliptic curve Curve25519 due to its speed. This means that each RIDER
uses the public key of the server, as well a unique public/private key to encrypt and transmit data. Lighthouse is
written in Python and depends on libzmq/libsodium.

\subsection{Homebase}

Homebase is a script that receives, decrypts and records all information received from Lighthouse and stores them in the
RIDER database. This allows remote monitoring of drive space and system health. All RIDER key management is done here in
order to decrypt messages from each unique box.

\subsection{Heartbeat}

Heartbeat is an engineer facing interface that displays RIDER system status information in order to validate successful
operation or gain insights as to potential system malfunction.  Heartbeat uses the information committed to the database
from Homebase to keep track of various RIDER logs. This is useful for analyzing the current state of the vehicle fleet,
and assists in determining which instrumented vehicles are in need of drive swaps (due to the hard drive running out of
space) or system repairs. It is also useful for verifying that any repairs made were successful.

\subsection{RIDER Database}

A PostgreSQL database is used to store all incoming trip information, as well as to house information about all trips offloaded
to a storage server. After additional processing, useful information about each trip can be added to the
database. Queries can then be structured to obtain specific trips or times in which specific events or conditions
occurred. The following tables are fundamental to the trip processing pipeline:
\begin{itemize}
  \item \textbf{instrumentations}: dates and vehicle IDs for the installation of RIDER boxes
	\item \textbf{participations}: unique subject and study IDs are combined to identify primary and secondary drivers
	\item \textbf{riders}: rider IDs paired with notes and IP addresses
  \item \textbf{vehicles}: vehicle information is paired with vehicle IDs such as the make and model, the manufacture date, color, 
  and availability of specific technologies
  \item \textbf{trips}: provides a unique ID for each centrally offloaded trip as 
  well as the study, vehicle, subject and rider IDs. Also provides information about
  synchronization state, available camera types and subsystem data. Metadata about 
  the content of the trip itself is included, such as the presence of sun, gps 
  frequency and the presence of certain technology uses or acceleration events.
  \item \textbf{epochs_\textit{epoch-label}}: tables for each epoch type are labeled and used to identify trips and 
  video frame ranges for which they occur (e.g. autopilot use in Teslas would be in epochs_autopilot)
  \item \textbf{homebase_log}: contains streamed log information from the homebase 
  script that keeps track of RIDER system health and state
\end{itemize}

\subsection{Cleaning} 
After raw trip data is offloaded to a storage server, all trips must be inspected for any inconsistencies. 
Some trips may have inconsistencies that can be fixed, as in the case where timestamping information could be
obtained from multiple files, or when a nonessential subsystem failed during a trip (e.g. IMU or audio).
In unrecoverable cases, like the event where a camera was unplugged during a trip, that trip is removed from the dataset.
Trips that have valid data files may also be removed from the dataset if that trip meets some set of filtering constraints,
like when a vehicle is turned on, but does not move before turning off again.

\subsection{Synchronization}
%\lexdo{Check, modify, or remove the summarized synchronization process below}

After completing cleaning and filtration, valid trips undergo a series of synchronization steps.  First, the timestamps
of every frame gathered from every camera are aligned in a single video CSV file at 30 frames per second using the
latest camera start timestamp and the earliest camera end timestamp. In low lighting conditions the cameras may drop to
recording at 15 frames per second. In these cases, some frames may be repeated to achieve 30 frames per second in the
synced video.

After all raw videos have been aligned, new synchronized video files can then be created at 30 frames per second.  CAN
data is then decoded by creating a CSV with all relevant CAN messages as columns and synced frame IDs as rows.  CAN
message values are then inserted frame-by-frame based on the closest timestamp to each decoded CAN message.  A final
synchronized visualization can then be generated that shows all video streams and CAN info in separate panels in the
same video.  The data is then ready to be processed by any algorithm running statistics, detection tasks, or manual
annotation tasks.

\section{Trips and Files}

This section will define how trip data files may be stored in a trip directory. A trip directory represents a trip that
a driver took with their vehicle from start to finish.  These are the files that are offloaded from the external storage
drive in a RIDER box onto a central server, where the data can be cleaned, synchronized, or processed in some other way.

\subsection{Trip Configuration Files}

Trip configuration files store specifications and information about available subsystems are included to manage the data logging process.

\begin{itemize}
\item \textbf{trip_dacman.json}: a configuration file containing subject and systems information used to record the trip
\item \textbf{trip_diagnostics.log}: a text file containing diagnostics information recorded during the trip: includes external
  temperature, PMU temperature, HDD temperature, power usage and free disk space
\item \textbf{trip_specs.json}: a json file containing start and end timestamps for all subsystems
\end{itemize}
 
\subsection{Trip Data Files}

Trip data files are the end point of all recording RIDER data streams. They include numerous CSV (comma separated values) files
that provide timestamping information, as well as raw video files in H264 and audio files in RAW formats. 

\begin{itemize}
\item \textbf{\textit{camera-directory}}: a directory named by camera type (all contained files are also named by that camera type)
\begin{itemize}
\item \textbf{\textit{camera-name}.h264}: a raw H264 file 
\item \textbf{\textit{camera-name}.error}: contains camera-specific errors
\item \textbf{\textit{camera-name}.csv}: matches recorded frames with system timestamps for later synchronization
  \begin{itemize}
    \item \texttt{frame,ts_micro}
  \end{itemize}
\end{itemize}
\item \textbf{data_can.csv}: contains CAN data
  \begin{itemize}
    \item \texttt{ts_micro, arbitration_id, data_length, packet_data}
  \end{itemize}
\item \textbf{data_gps.csv}: contains GPS data
  \begin{itemize}
    \item \texttt{ts_micro, latitude, longitude, altitude, speed, track, climb}
  \end{itemize} 
\item \textbf{data_imu.csv}: contains IMU data
  \begin{itemize} 
    \item \texttt{ts_micro, x_accel, y_accel, z_accel, roll, pitch, yaw}
  \end{itemize}
\item \textbf{audio.raw}: contains raw output from a specified camera
\item \textbf{can.error, gps.error, imu.error, audio.error}: text-based error files for CAN, GPS, IMU and audio recordings
\end{itemize}

\subsection{Cleaning Criteria}

The following cases represent recoverable errors that a trip may contain, as well as their implemented solutions:
\begin{itemize}
  \item \textbf{Invalid permissions}: UNIX permissions of the trip directory must allow group-only read/write access
  \item \textbf{Missing backup}: Raw essential files are backed up to allow a rollback to previous versions
  \item \textbf{Missing trip_specs.json}: The trip_specs.json file can sometimes be reconstructed using recorded timestamps
  \item \textbf{Missing or invalid ID}: Vehicle, camera or subject IDs may be corrected based on trip context
  \item \textbf{Invalid Nonessential Files}: If IMU or audio have failed, they can be removed and the trip can be preserved
  \item \textbf{Invalid last CSV line}: Interrupted subsystems may write incomplete lines to their data file, which can be removed
\end{itemize}

\subsection{Filtering Criteria}

The following cases represent unrecoverable errors or chosen criteria that result in the removal of a trip from the dataset:
\begin{itemize}
\item \textbf{Nonconsenting driver}: When the driver is not a consented participant in the study
\item \textbf{Requested removal}: When the subject requests certain trips, dates or times be removed
\item \textbf{Vehicle doesn't move}: When the kinematics of the vehicle indicate no change in speed
\item \textbf{Trip data files $<$ 15MB}: When the total size of a trip's files are less than 15MB (faster than duration checks)
\item \textbf{Trip duration $<$ 30 seconds}: When the shortest camera recording is less than 30 seconds in duration
\item \textbf{Missing essential files}: When camera files, \texttt{trip_dacman.json} or \texttt{data_can.csv} are missing
\item \textbf{Outside volunteer participation range}: Indicative of MIT staff driving the vehicle to be maintained or washed
\item \textbf{Large essential subsystem error files}: When there are many errors for a camera or for CAN
\item \textbf{Mismatches in subsystem timestamps}: When one subsystem ends at least one minute earlier than another
\end{itemize}

\subsection{Synchronized Files}

%\lexdo{Decide if this section, along with the earlier summary of synchronization, is worth including}

Synchronized files are created by synchronization scripts that run after cleaning and filtering has taken place. These scripts 
align video frames and CAN messages at a rate of 30 frames per second. They are created using the same trip naming convention in a separate, 
processed directory.

\begin{itemize}
  \item \textbf{synced_video.csv}: every row contains a video frame ID and timestamp from every camera at 30 frames per second  
  \item \textbf{synced_video_\textit{camera-name}.mp4}: Synchronized with all other videos at 30 FPS using H264 encoding
  \item \textbf{synced_can.csv}: each row represents a synced video frame and the closest CAN values associated with that timestamp for every CAN message
  \item \textbf{synced_vis_panels.mp4}: an optional visualization video file that displays all synced videos in separate panels where CAN data may be also displayed
\end{itemize}

\section{Ongoing Hardware Development and Innovation}

RIDER is an instrumentation platform that has been proven through extensive testing to have adequate data collection
abilities for naturalistic driving research. During the research, development, and testing process we met some
limitations of the system. While a single board computer is sufficient for most collection processes, limitations of
minimal system memory could create issues when expanding the system. Similarly, a Dual-Core ARM processor is very
capable when interfacing with sensors and writing data out to files, but performance can fluctuate if any preprocessing
of the data is required onboard. From our work we have proposed the following improvements to some of these common
issues.

The largest enhancement for the entire RIDER system would be to upgrade the single board computing platform. Use of the
NVIDIA Jetson TX2 would provide more expandability both for I/O and processing. With greater processing and GPU
bandwidth available, real-time systems could be implemented using both video and sensor data simultaneously for
detection and driver warning systems, internal annotation of data and more. With greater I/O capability, upgraded
sensors packages with higher data bandwidths can be implemented. Much like the Banana Pi Pro the Jetson TX2 has not one,
but two fully supported CAN controllers to interface with a secondary CANbus system on the vehicle. Jetson TX2 has
expandability not only for SATA but also PCIe and mSATA, allowing for even greater expansion of third party modules. The
enhanced processing via CPU and GPU with 8 times the onboard RAM allows the potential for preprocessing and integration
of real-time driver monitoring systems. The Jetson also has the major advantage of being supported for use in multiple
configurations for in vehicle applications.

\section{Conclusion}\label{sec:conclusion}

% The application of state-of-the-art embedded system programming, software engineering, data processing, distributed
% computing, computer vision and deep learning techniques to the collection and analysis of large-scale naturalistic
% driving data in the MIT-AVT study seeks to break new ground in offering insights into how human and autonomous vehicles
% interact in the rapidly changing transportation system. This work presents the methodology behind the MIT-AVT study
% which aims to define and inspire the next generation of naturalistic driving studies.
% % The governing design principle of
% % this study is that, in addition to prior successful NDS approaches, we leverage the power of computer vision and deep
% % learning to automatically extract patterns of human interaction with various levels of autonomous vehicle technology. We
% % both (1) use AI to analyze the entirety of the driving experience in large-scale data and (2) use human expertise and
% % qualitative analysis to dive deep into the data to gain case-specific understanding.
% To date, the dataset includes \subjects{} participants, \days{} days of participation, \miles{} miles, and \frames{}
% video frames. Statistics about the size and scope of the MIT-AVT dataset are updated regularly on
% \url{https://hcai.mit.edu/avt}.

The application of state-of-the-art embedded system programming, software engineering, data processing, distributed
computing, computer vision and deep learning techniques to the collection and analysis of large-scale naturalistic
driving data in the MIT-AVT study seeks to break new ground in offering insights into how human and autonomous vehicles
interact in the rapidly changing transportation system. This work presents the methodology behind the MIT-AVT study
which aims to define and inspire the next generation of naturalistic driving studies. To date, the dataset includes
\subjects{} participants, \days{} days of participation, \miles{} miles, and \frames{} video frames. Statistics about
the size and scope of the MIT-AVT dataset are updated regularly on \url{https://hcai.mit.edu/avt}.

\section*{Acknowledgment}\label{sec:acknowledgement}

The authors would like to thank MIT colleagues and the broader driving and artificial intelligence research community
for their valuable feedback and discussions throughout the development and on-going operation of this study, especially
Joseph F. Coughlin, Sertac Karaman, William T. Freeman, John Leonard, Ruth Rosenholtz, Karl Iagnemma, and all the
members of the AVT consortium.

The authors would also like to thank the many vehicle owners who have provided and continue to provide valuable insights
(via email or in-person discussion) about their experiences interacting with these systems. Lastly, the authors would
like to thank the annotation teams at MIT and Touchstone Evaluations for their help in continually evolving a
state-of-the-art framework for annotation and discovering new essential elements necessary for understanding human
behavior in the context of advanced vehicle technologies.

Support for this work was provided by the Advanced Vehicle Technology (AVT) consortium at MIT. The views and conclusions
being expressed are those of the authors, and have not been sponsored, approved, or necessarily endorsed by members of
the consortium. All authors listed as affiliated with MIT contributed to the work only during their time at MIT as
employees or visiting graduate students.

\balance

\bibliographystyle{IEEEtran}
\bibliography{avt}

\end{document}